# PERCEPTUAL COPYRIGHT PROTECTION USING MULTIRESOLUTION WAVELET-BASED WATERMARKING AND FUZZY LOGIC

Ming-Shing Hsieh

Department of Computer Engineering and Information Science,
Aletheia University, Damshui, Taiwan 251
sms.sms@msa.hinet.net

## ABSTRACT

*In this paper, an efficiently DWT-based watermarking technique is proposed to embed signatures in images to attest the owner identification and discourage the unauthorized copying. This paper deals with a fuzzy inference filter to choose the larger entropy of coefficients to embed watermarks. Unlike most previous watermarking frameworks which embedded watermarks in the larger coefficients of inner coarser subbands, the proposed technique is based on utilizing a context model and fuzzy inference filter by embedding watermarks in the larger-entropy coefficients of coarser DWT subbands. The proposed approaches allow us to embed adaptive casting degree of watermarks for transparency and robustness to the general image-processing attacks such as smoothing, sharpening, and JPEG compression. The approach has no need the original host image to extract watermarks. Our schemes have been shown to provide very good results in both image transparency and robustness.*

## 1. INTRODUCTION

The increasingly easy access to digital media and increasingly powerful tools available for manipulating digital media have made media security a very important issues. Due to the open environment of Internet downloading, copyright protection introduces a new set of challenging problems regarding security and illegal distribution of privately owned images. One potential solution for declaring the ownership of the images is to use *watermarks* to embed an invisible signal into multimedia data so as to prove the owner identification of the data and discourage the unauthorized copying. In the study, we focus our research on digital image watermarking, but the method could be modified for other applications as well.

In general, there are two common requirements of watermark. First requirement is the watermarks must be perceptually transparency. They should not noticeable to the viewer. The second requirement is the watermarks must be robust to intentional or unintentional attacks and common signal processing. Particularly, the watermark should still be detectable even after attacks have been applied to the watermarked image. These attacks include image compression, linear or nonlinear filtering, image enhancements, etc.

The watermarking techniques can be divided into two different classifications. One is applied to spatial domain, and the other is applied to the frequency domain. The spatial domain watermark techniques are developed early [1, 2], simple but not robust is their obvious weakness. They can't against intentional or unintentional attacks and image processing. The embedded watermark signals are easily disfigured, distorted, or removed. The frequency domain approach has some advantages because most of the signal processing operations can be well characterized, and many good perceptual models are developed in the frequency domain [3].

Cox *et al.* [4] proposed an image watermarking method based on spread spectrum theory, which shows good performance among invisibility and robustness to signal processing operation and common geometric transform. Hsu *et al.* [5] proposed *DCT* based image watermarking technique by embedding a visually recognizable watermark. An image-content-based adaptive embedding scheme is applied in discrete Fourier transform (*DFT*) domain of each perceptually high textured subimage to ensure better visual quality and more robustness [6]. On the other hand, several methods used the discrete wavelet transform (*DWT*) to hide data to the frequency domain to provide extra robustness against attacks.

*DWT* is known that the wavelet image/video coding, such as embedded zero-tree wavelet (*EZW*) coding [7], and play an important role in image/video compression standards, such as *JPEG2000* and *MPEG4* due to its excellent performance in compression. In [8], they propose watermarking schemes where visual models are used to determine image dependent upper bounds on watermark insertion. This allows users to provide the maximum strength transparent watermark, which is extremely robust to common image processing and editing such as image compression. Dugad *et al.* [9] picked all coefficients in the subbands, which are larger than a given threshold, and watermark is added to these coefficients only. The reason to set the threshold is to embed the watermarks in the edge regions of the host image, so that the watermarked image is not distinguishable from the original image. Hsu *et al.* [3] used *DWT* other than *DCT* and the same method proposed in [5] to embed signal into each wavelet subbands. The empirical results showed that the watermarked image was more robustness and imperceptibly then they early did in [3]. Inoue *et al.* [10] proposed a method based on *zerotree* [11], which classified wavelet coefficients as insignificant or significant using *zerotree*, and select proper position to insert watermarks. Embedding watermarks into host image using *DWT* are both in transparency and robustness.

In [4] the watermark is a symbol or a random number, which comprises of a sequence of bits, and can only be detected by employing detection theory. In [12], [13], the watermark is a visually recognizable pattern. This kind of watermark is more intuitive for representing one's identity than a random sequence, and also can be measured the correlation between the detected visual pattern and original watermark during the verification phase.

Originally, the determination of a wavelet coefficient to be an embedded target is done by comparing subband's coefficients uniformly with a fixed sorting order. If all coefficients of a wavelet subband are less than the threshold value, they are set "insignificant" and had no chance to embed watermarks. Actually, the determination is uncertain and mutual influence. For example, all coefficients of a subband being slightly less than the threshold value should be not always set "insignificant"; on the other hand, all coefficients of a subband being slightly greater than the threshold value should be not always set "significant". Therefore, a fuzzy filter is necessary to provide reasoning with the vague and uncertain information. The use of fuzzy filters in image processing is based on the idea that pixels or coefficients are not uniformly fired by every fuzzy rule. The fuzzy set theory has the potential capability to efficiently represent input and output relationships of dynamic systems, thus it has gained popularity. For example, the usage of fuzzy algorithms for vector quantization has been proposed by Munteanu *et al.* [14]; fuzzy clustering was employed to preserve the textually important image characteristics while a compression algorithm was proposed by Karras *et al.* [15]; Yang and Toh [16] applied heuristic fuzzy rules to improve the performance of the traditional multilevel filter; Farbiz *et al.* [17] applied a new fuzzy logic filter to improve the performance of image enhancement. Hsieh *et al.* [18] applied a new fuzzy logic filter to improve the performance of image coding. In [19], an adaptive watermarking algorithm is presented which exploits a biorthogonal wavelets-based human visual system (*HVS*) and a Fuzzy Inference System (*FIS*) to protect the copyright of images in learning object repositories. The *FIS* isutilized to compute the optimum watermark weighting function thatwould enable the embedding of the maximum-energy and imperceptiblewatermark.

In this paper, we propose a wavelet-based watermarking approach by adding visually recognizable image to the larger entropies of coefficients calculated by the fuzzy inference filter of selected wavelet subband. The proposed approach has the following advantages: (i) the extracted watermark is visually recognizable to claim one's ownership, (ii) the approach is hierarchical and has multiresolution characteristics, (iii) the embedded watermark is hard detected by human visual perceptivity. Our experimental results show that the proposed watermarking approach is very robust to image compression and image operations.

This paper is organized as follows. Wavelet transform of images, context-based method and fuzzy filter is described in Section 2. Section 3 describes the watermark embedding approach. In Section 4, the experimental results are shown. The conclusion of this paper is stated in Section 5.

## 2. PRELIMINARIES

### 2.1. Wavelet transform of images

The wavelet transform is identical to a hierarchical subband system, where the subbands are logarithmically spaced in frequency. For an input sequence of length $N$, *DWT* will generate an output sequence of length $N$. The 1-D *DWT* can be implemented by the *Direct Pyramid Algorithm*, which was developed by Mallat [20] as follows.

Begin {*Direct Pyramid Algorithm*}

For $k = 1$ to $K$

For $n = 1$ to $2^{K-k}$

$$Y[k, n] = \sum_{m=0}^{N-1} X[k-1,\ 2n-m]\ \ g(n)$$

$$X[k, n] = \sum_{m=0}^{N-1} X[k-1,\ 2n-m]\ \ h(n)$$

End

where $k$ is the current octave, $K$ represents the total number of octaves, $1 \le k \le K$, $n$ is the current input, and $N$ is the total number of inputs, $1 \le n \le N$. Mallat's direct pyramid algorithm is executed by down sampling.

For a 2-D image, a wavelet $\Psi$ and a scaling function $\Phi$ are chosen such that the scaling function $\Phi_{LL}(x, y)$ of low-low subband in a 2-D wavelet transform can be written as $\Phi_{LL}(x, y) = \Phi(x)\ \Phi(y)$. Three other bi-dimensional wavelets can also be obtained using the wavelet associated function $\Psi(x)$ as follows.

$\Psi_{LH}(x, y) = \Phi(x)\ \Psi(y)$ ; horizontal

$\Psi_{HL}(x, y) = \Psi(x)\ \Phi(y)$ ; vertical

$\Psi_{HH}(x, y) = \Psi(x)\ \Psi(y)$ ; diagonal

where $H$ is a high-pass filter and $L$ is a low-pass filter.

The basic idea in the *DWT* of a 2-D image is as follows. An image is firstly decomposed into four parts of high, middle, and low frequencies (i.e., $LL_1$, $HL_1$, $LH_1$, $HH_1$) subbands, by cascading horizontal and vertical two-channel critically subsampled filter banks. The subbands labeled $HL_1$, $LH_1$, and $HH_1$ represent the finest scale wavelet coefficients. To obtain the next coarser scale of wavelet coefficients, the subband $LL_1$ is further decomposed and critically subsampled. This process is continued an arbitrary number of times, which is determined by the application at hand. Fig. 1 shows layout of the image subbands from three-level dyadic decomposition and an example of *DWT* decomposition of the *Lena* image using a wavelet filter set. In the figure, *Lena* image is decomposed into ten subbands for three scales. Each level has various band-information such as low-low, low-high, high-low, and high-high frequency bands. Furthermore, from these *DWT* coefficients, the original image can be reconstructed. The reconstruction process is called the inverse *DWT* (*IDWT*). Let $I\ [m, n]$ represent an image. The *DWT* and *IDWT* for $I\ [m, n]$ can be similarly defined by implementing the *DWT* and *IDWT* for each dimension *m* and *n* separately: $DWT_n\,[DWT_m\, I\,[m, n]]$.

The wavelet transform is identical to a hierarchical subband system, where the subbands are logarithmically spaced in frequency and represent an octaveband decomposition. An example of three scales wavelet transform is show in Fig. 1. The image is first decomposed into for subband $LL_1$, $LH_1$, $HL_1$, $HH_1$ by using horizontal and vertical two channel critically subsampled filter banks. Each coefficient represents a spatial area corresponding to approximately a 2×2 area of the original image. The $LH_1$, $HL_1$, $HH_1$ represent the finest scale wavelet coefficient. After decomposed and critically subsampled the subband $LL_1$, we can obtain the next coarser scale of wavelet coefficients. Note that for each coarser scale, the coefficients represent a larger spatial area of the image but a narrower band of frequencies. The decomposed process continues until some final scale is reach.

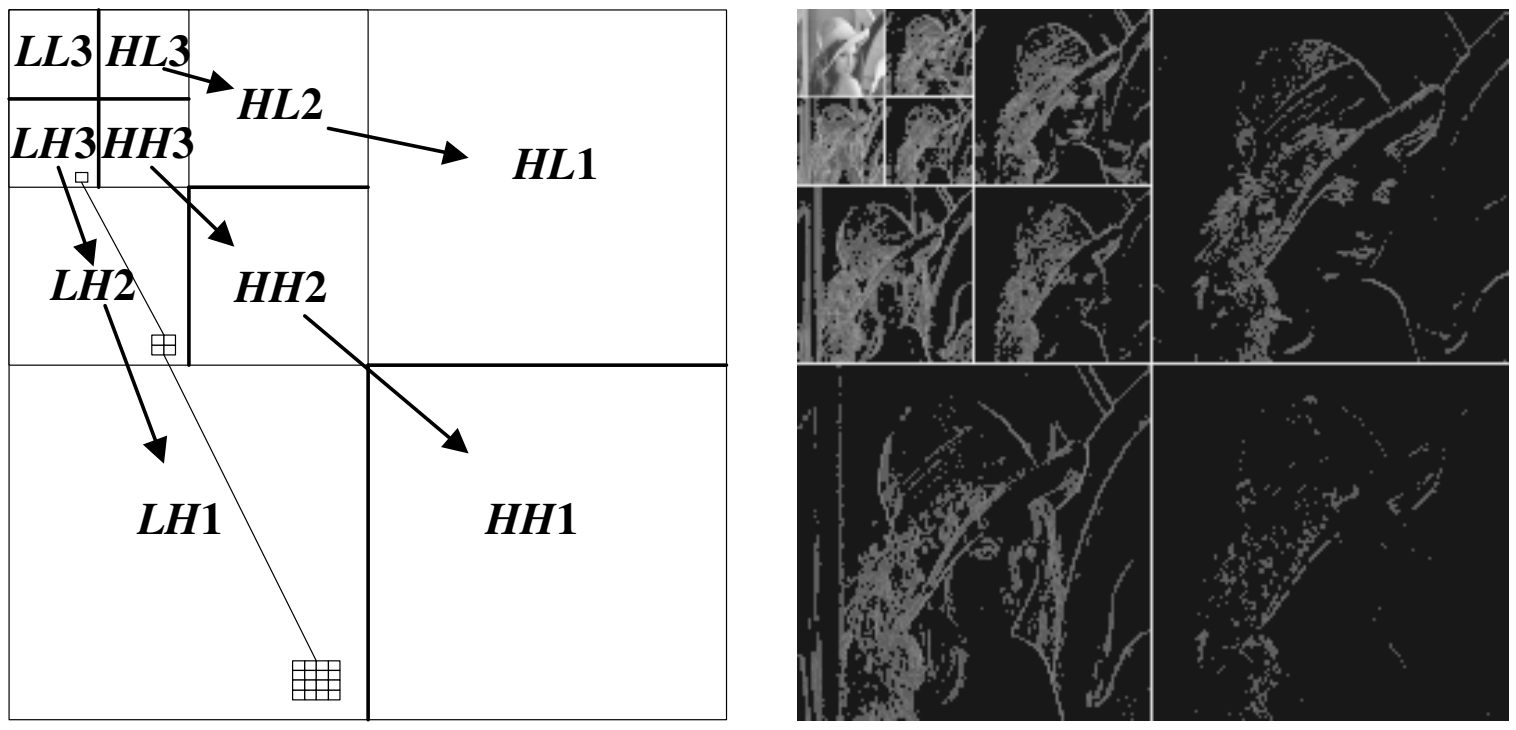

(a) (b)

Fig. 1. (a) Layout of the image subbands from three-level dyadic decomposition. (b) An example of *DWT* decomposition of the *Lena* image.

## 2.2. Context-based method

The main function of the context-based method is to find the relationship of one pixel/coefficient and its adjacent pixels/coefficients in the image. The relationship among the context is to determine characteristics of the target pixel/coefficient. Context-based method can be applied in image compression, which is based on wavelet transform. Triantafyllidis *et al.* [21] proposed that current wavelet coefficient could be estimated using the coefficients that lie in the current band, scale bound(s), parent band, and the pyramid structure. Yoo *et al.* [21] introduce a context-based classification technique, which classifies each subband coefficient based on the surrounding coefficients, and different quantizer is then used for each class. In our approaches, we introduce context-based method on watermark embedding. Unlike previous works that only use the larger coefficients to insert watermarks, we use the current coefficient and its surrounding coefficients to calculate the entropy of each coefficient in selected subband, and choose the larger entropy to embed watermarks. If a coefficient has higher entropy denotes the violent variation of spatial domain in the host image, to embed watermarks in the center coefficient of the context could improve the transparency of the watermarked image and robustness to extracted watermarks.

## 2.3. Using fuzzy filter to calculate entropy of wavelet coefficient

In this section, the use of context and fuzzy filter to calculate the entropy corresponding to each coefficient in one subband are described. General block diagram of a fuzzy inference system is shown in Fig. 2.

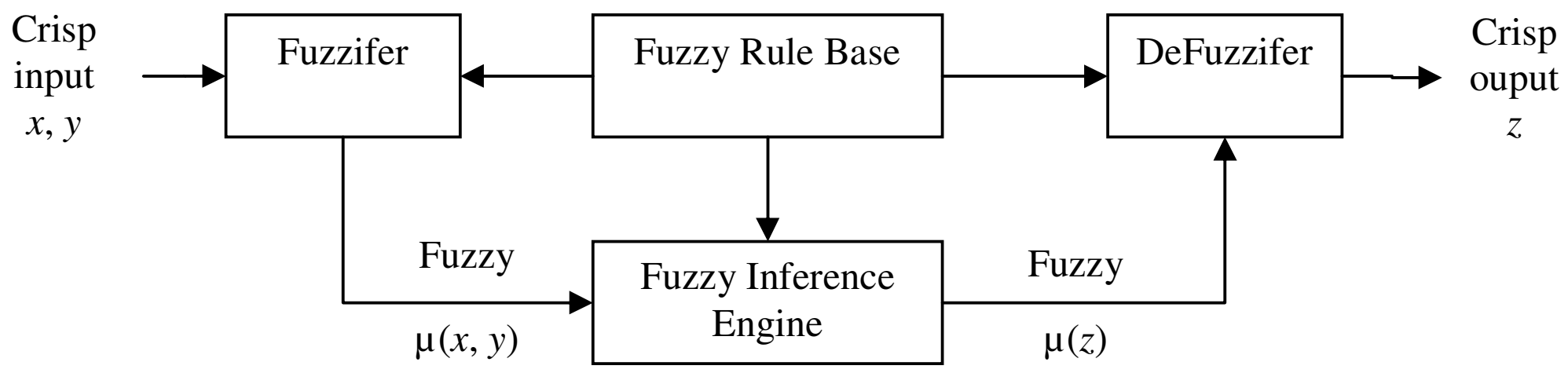

Fig. 2. General block diagram of a fuzzy inference system (*FIS*).

In an arbitrary wavelet subband *S*, where we determine entropy of each coefficient in *S* is based on current coefficient and its surrounding coefficients. Template made of nine coefficients form the context is shown in Fig. 3.

Let $x_0$ be the current target coefficient to estimate its entropy, $x_i$, $1 \le i \le 8$, is a $x_0$'s surrounding coefficient as shown in Fig. 3. Fuzzy rules shown in Table 1 are used as inference filter to calculate the entropy corresponding to each coefficient in *S*:

| | | |
|---|---|---|
| $x_1$ | $x_2$ | $x_3$ |
| $x_4$ | $x_0$ | $x_5$ |
| $x_6$ | $x_7$ | $x_8$ |

Fig. 3. Template made of nine coefficients form the context.

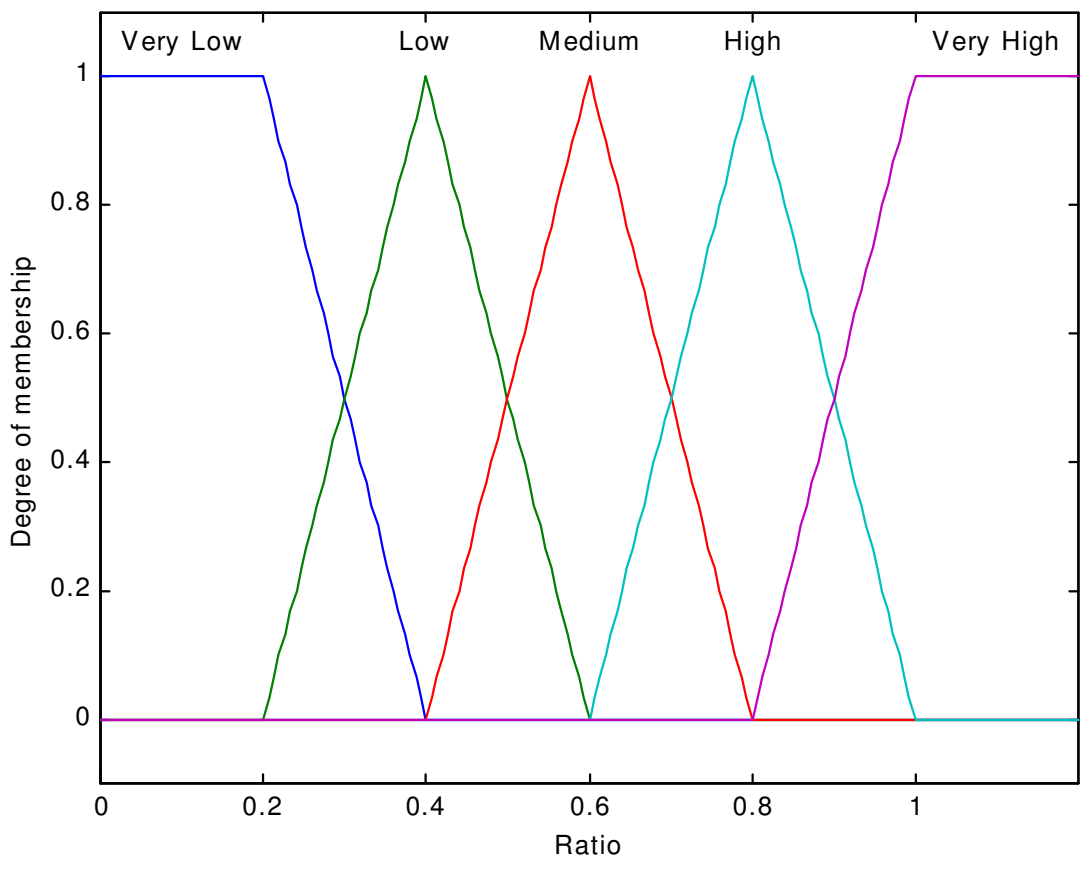

Fig. 4. Membership functions depicting the degrees of *normalized fuzzy wavelet coefficients*.

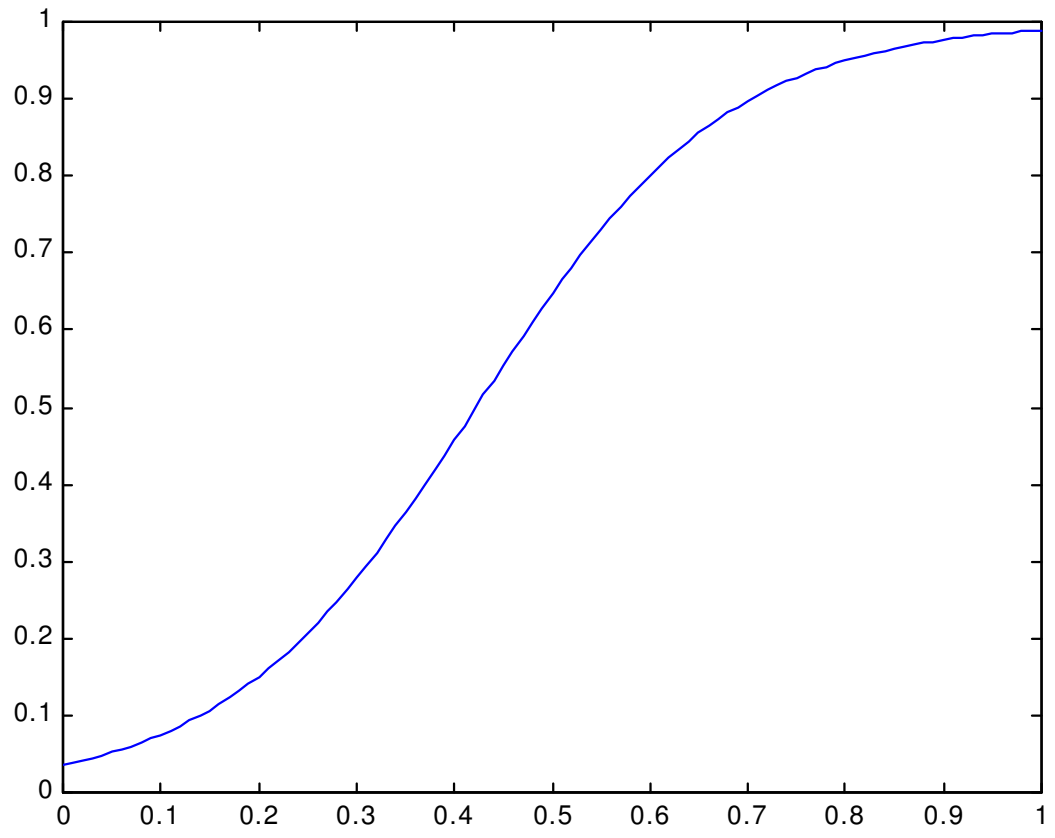

Fig. 5. Typical membership function for the fuzzy function "more".

Table 1. Fuzzy Inference rules

| Rule | Rule Description |
|---|---|
| $R_1$ | $R_1$: If more of $x_i$ are *Very High* Then $En(x_i)$ is *Very High*. |
| $R_2$ | $R_2$: If more of $x_i$ are *High* Then $En(x_i)$ is *High*. |
| $R_3$ | $R_3$: If more of $x_i$ are *Medium* Then $En(x_i)$ is *Medium*. |
| $R_4$ | $R_4$: If more of $x_i$ are *Low* Then $En(x_i)$ is *Low*. |
| $R_5$ | $R_5$: If more of $x_i$ are *Very Low* Then $En(x_i)$ is *Very Low*. |

Note that in the above production rules in Table 1, $x_i$'s, named *normalized fuzzy coefficients* (*NFCs*), are the *normalized fuzzy value* (*NFV*) of wavelet coefficients. *NFV* of each coefficient is normalized by dividing some of $T$ to emphasis the importance of coefficients in its context. The predicate's and conclusion's part, '*Very High*', '*High*', '*Medium*', '*Low*', and '*Very Low*' are the degrees of fuzzy membership as illustrated in Fig. 4. The output $En(x_i)$ denotes the entropy of $x_i$, is given by the target coefficient and its surrounding coefficients which yield the result by *Equ* (3). The term *more* denotes an *S*-type fuzzy function whose typical shape is shown in Fig. 5. The curve of this function enables the non-uniform firing of the basic fuzzy rules. This *more* function may be described by the following formula:

$$\mu_{more}(z) = \frac{1}{1+e^{-(\alpha z-\beta)}} \tag{1}$$

where $\alpha$=7.80375, $\beta$=3.29596. The curve of the function enables the non-uniform firing of the basic fuzzy rules.

*NFV* of each coefficient is defined as follows.

$$x_0 = \frac{x_0}{T_0},\ x_{n|n=2,4,5,7} = \frac{x_n}{T_1},\ x_{m|m=1,3,6,8} = \frac{x_m}{T_2},$$

where $T_0 = Avg(|S|)$, $T_1 = T_0+1/16\times Std(|S|)$, $T_2 = T_0+1/8\times Std(|S|)$, *Avg*, *Std* denote the average, stand deviation, respectively.

The activity degree of rule $R_k$ is computed by the following relationship:

$$\lambda_k = \min\{\mu_{quality}(x_i) : x_i \in \sup(quality)\}\times \mu_{more}\left[\frac{number\ of\ x_i\ \ which\ x_i \in \sup(quality)}{total\ number\ of\ x_i}\right] \tag{2}$$

where $1\leq k \leq 5$, and the corresponding quality $\in$ {*Very High* | *High* | *Medium* | *Low*}.

After all $\lambda_k$'s are calculated, the entropy of coefficient $x_i$ is computed by equation:

$$En(x_i) = \sum_{k=1}^{5} C_k \lambda_k \tag{3}$$

where $C_k$ represents the center point of the membership function $R_k$.

## 3. WATERMARKING IN THE DWT DOMAIN

The proposed digital watermarking approach can hide visually recognizable patterns in images. The main study task of digital watermarking is to make watermarks invisible to human eyes as well as robust to various attacks. In the proposed approach, watermarks are embedded in the host image by modifying the coefficients with larger weighted energy. The block diagram of embedding watermarks in the host image is shown in Fig. 6.

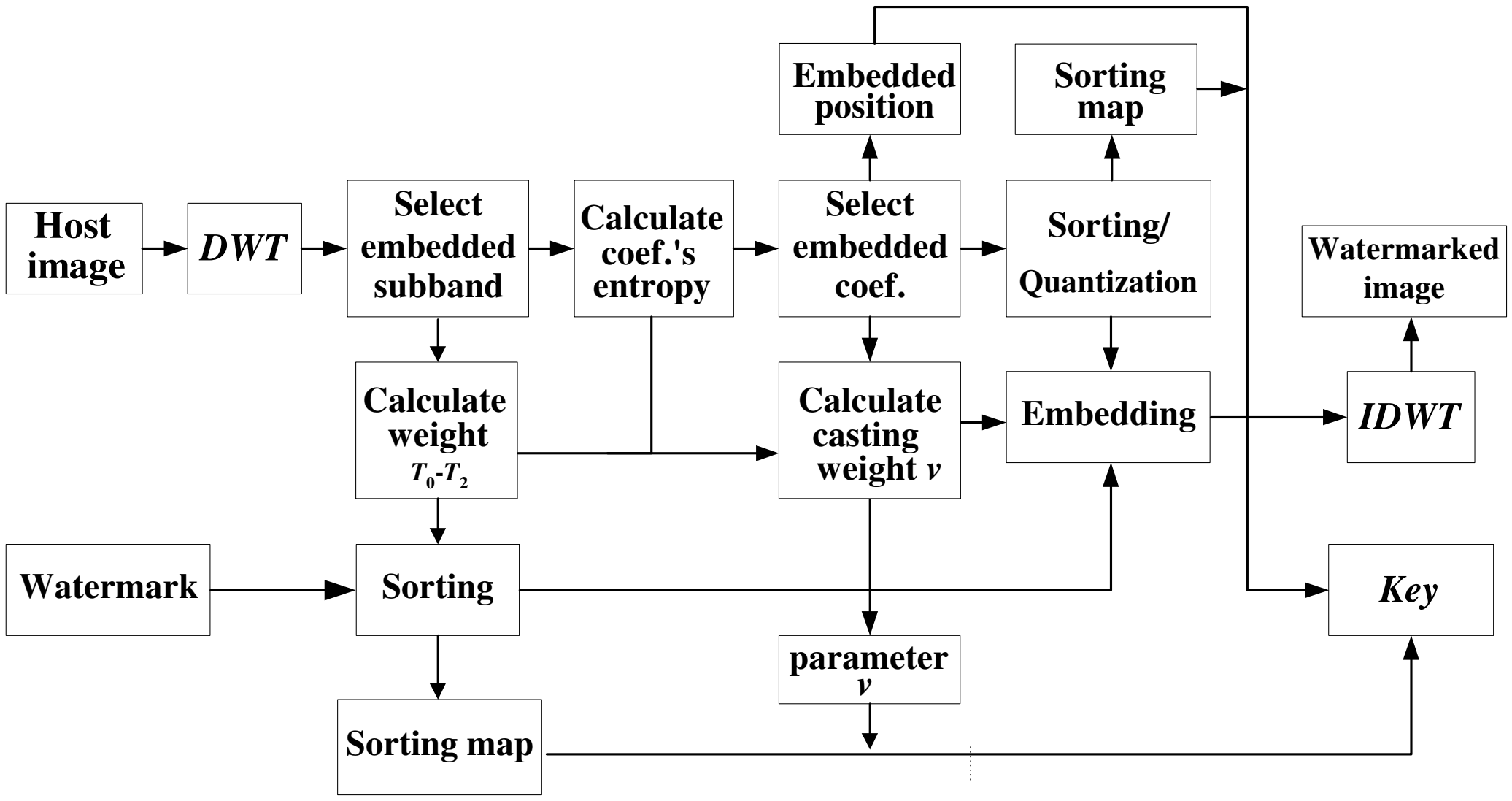

Fig. 6. The block diagram of embedding watermarks.

## 3.1. Watermark embedding method

The algorithm for embedding a watermark in a host image is described as follows:

*Step* 1: Sort the gray levels of watermark $W$ in ascending order to generate the sorted watermark $SW$, Assume that the size of watermark image $W$ is $n$.

*Step* 2: Decompose a host image $H$ into three levels with ten subbands of a wavelet pyramid structure. Choose a subband $S$, for example $HL_3$, to embed watermark $W$.

*Step* 3: Calculate the weighted entropy $E_n$ of coefficients in subband $HL_3$ using the proposed method approach described in Section II.*B*.

*Step* 4: Let the preset interval be $\tau$, $t$ be the number of referenced coefficients used as a key to extract watermark $W$ without the host image. $p$ coefficients with larger entropy are chosen from subband $S$, where $p = n + t$. The larger entropy of coefficients makes the watermarked image more robustness and transparency. Obviously, if $n/\tau$=0, the value of $t$ is $fix(n/\tau) + 1$, otherwise $t$ is $fix(n/\tau) + 2$, where function $fix$ is to get the integer part of its argument. Let $\{Cp\}$ be the set of referenced coefficients and the coefficients to be embedded watermarks; $\{Cp\}$ is called *the alternative coefficients*. Sorting $\{Cp\}$ to generate $\{SCp\}$ called the *sorted alternative coefficients*.

*Step* 4: Quantize $\{SCp\}$ using a preset interval could extract watermark $W$ without the host image.

*Step* 6: Embed watermark $SW$ into subband $HL_3$ by the watermark embedding strategy, $SC_j = SC_j + v_i \times SW_{index}$, where $SC_j$ is a *sorted alternative coefficients* except the referenced coefficients, $v_i = En_i \times (T_0+T_1+T_2)/3 = En_i \times T_1$. The value of scaling factor $v_i$ is the maximum variances to modify the embedded coefficients for robustness while embedding watermarks to the target coefficients and $SW_{index}$ is a sorted waremark. A coefficient with larger weighed entropy could embed watermark with larger scaling factor for robustness without obviously degrading the host image. The embedding procedure can be implemented by the proposed *Casting watermarks Algorithm* as follows.

*Begin* {*Casting Watermarks Algorithm*}

*index*=1

/* *Embedded watermarks for each intervals with* $\tau$ *coefficients.* */

*For i* = 1 to $p$-$\tau$ *step* $\tau$+1

*For j* = $i$+1 to $i$+$\tau$

$SC_j = SC_j + v_j \times SW_{index}$

*index* = *index* + 1

*EndFor*

*EndFor*

/* *Embedded watermarks for the rest alternative coefficients.* */

*If* ($n$ *mod* $\tau$) $\neq 0$

*For j* = $p$-($n$ *mod* $\tau$) to $p$-1

$SC_j = SC_j + v_j \times SW_{index}$

*index* = *index* + 1

*EndFor*

*EndIf*

*End* {*Casting Watermarks Algorithm*}

*Step* 7: Save scaling factor $v_i$, the symbol of embedded subband, symbol table of $SC_i$, corresponsive map of $C_i$ and $SC_i$, and corresponsive map of $W_i$ and $SW_i$. To form watermarked image, one can take the *IDWT* of the modified *DWT* coefficients and the unchanged *DWT* coefficients.

## 3.2. Watermark extracting method

The embedded watermark can be detected based on the stored parameters after the wavelet decomposition of the watermarked image as follows:

*Step* 1: Decompose a watermarked image into three levels with ten subbands using *DWT*.

*Step* 2: Restore the scaling factor $v_i$, the symbol of embedded subband, symbol map of $SC_i$, corresponsive map of $C_i$ and $SC_i$, and corresponsive map of $W_i$ and $SW_i$.

*Step* 3: Extract the sorted watermarks by the proposed *Extracting Watermarks Algorithm* as follows:

*Begin* {*Extracting Watermarks Algorithm*}

*index*=1

/* *Extracting watermarks for each intervals with* $\tau$ *coefficients.* */

*For i* = 1 to $p$-$\tau$ *step* ($\tau$+1)

*For j* = $i$+1 to $i$+$\tau$

$value = (|SC_i| + |SC_{i+\tau+1}|) / 2$

$SW_{index} = (|SC_j| - value) / v_j$

*index* = *index* + 1

*EndFor*

*EndFor*

/* *Extracting watermarks for the rest alternative coefficients.* */

*If* ($n$ *mod* $\tau$) $\neq 0$

*For j* = *p*-(*n mod τ*) to *p*-1

*value*=(|$SC_{p-(n \bmod \tau)-1}$+ |$SCx'_p$|) / 2

$SW_{index}$ = ($SCx'_j$ – *value*) / $v_j$

*index* = *index* + 1

*EndFor*

*EndIf*

*End* {*Extracting Watermarks Algorithm*}

*Step* 4: Rearranging watermarks from corresponsive map of $W_i$ and $SW_i$ to get the extracted watermarks *W*'.

In our scheme, the extracted watermark *W*' is a visually recognizable image. A subjective measurement based on the standard correlation coefficient,

$$correlation = \frac{\sum (W' - \overline{W}')(W - \overline{W})}{\sqrt{\sum (W' - \overline{W}')^2}\sqrt{\sum (W - \overline{W})^2}} \quad (4)$$

is used to evaluate the quality of the extracted watermarks by measuring the similarity of the original watermark *W* and extracted watermark *W*'. The value of *correlation* is between zero and one. A larger value of *correlation* represents more similarity of the original watermarks and extracted watermarks.

## 4. EXPERIMENTS

A 512×512 *Lena* image was taken as the host image to embed a 32×32 binary watermark image with "NCU CSIE" characters following a sequence of attacks. The proposed approach emphasizes the local characteristics in the experiments, we also examined the quality of watermarked images and detectability of watermarks.

### 4.1. Image quality

The proposed perceptual watermarking framework was implemented for evaluating both properties of transparency and robustness. The two conflicting characteristics are referred to the embedded positions and casting degree of watermarks. For the embedded positions, as we chose the larger entropy of coefficients to embedded watermarks, which stands for the violent variance of spatial domain in a host image, the transparency and robustness of host image could be hewn. For the casting strength, the larger casting degree makes it more robust to the watermarked image. The side effect is that the lower transparence makes the watermarks the perceptual break. In the experiments, an adaptive casting strength was proposed to adjust casting degree of the coefficients, which makes the balance of transparency and robustness. Fig. 7 shows an example of embedding results, where *Lena* is used as the test image, a binary image with "NCU CSIE" is used as the watermark. The performance of the proposed image watermarking approach was evaluated. The common-used 512×512 gray-scale *Lena* image was used as original image and the peak signal-to-noise ratio (*PSNR*) was used for objective comparison,

$$PSNR = 10\log_{10}\frac{255^2}{MSE}, \quad (5)$$

where *MSE* is the mean square error between a watermarked image and its original image.

Fig. 7 also shows the watermarked image, in the figure, the *PSNR* of the watermarked image is 45.68, correlations of the original watermarks and extracted watermarks is 1.

Table 2 shows an example of extracting results from Fig. 7(b) without any attacks using the proposed method.

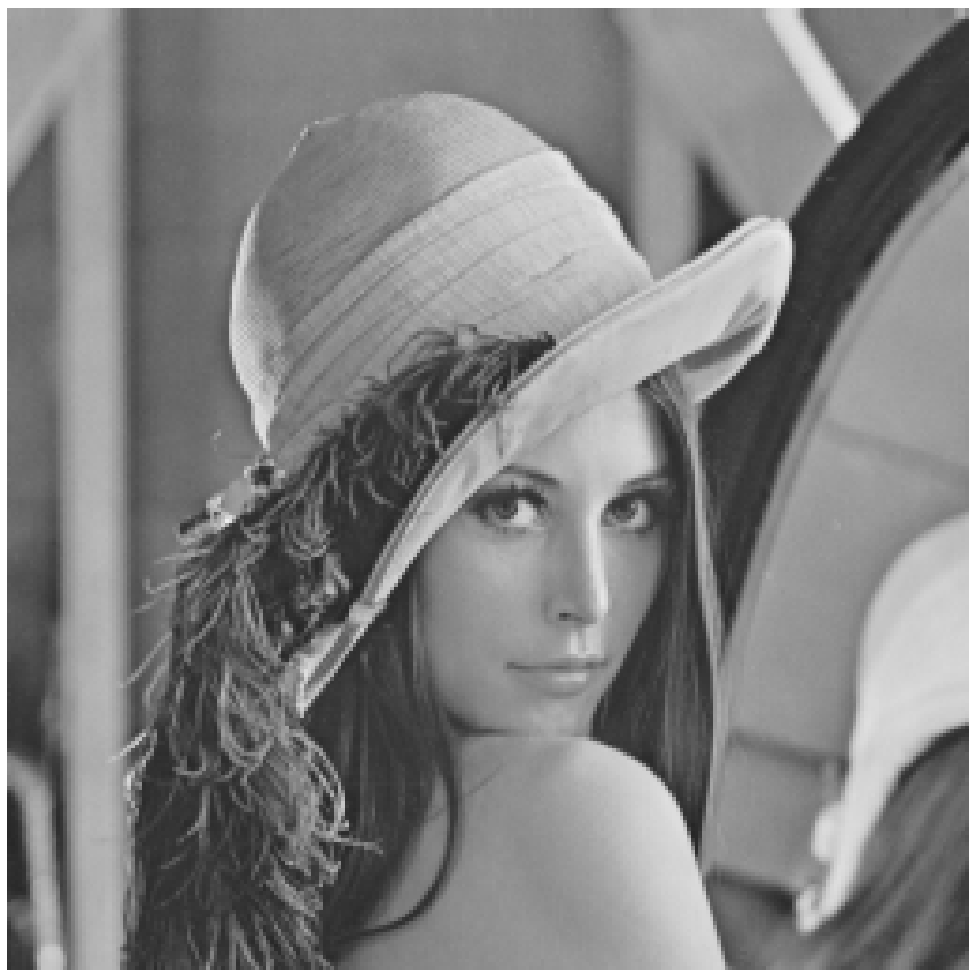

(a)

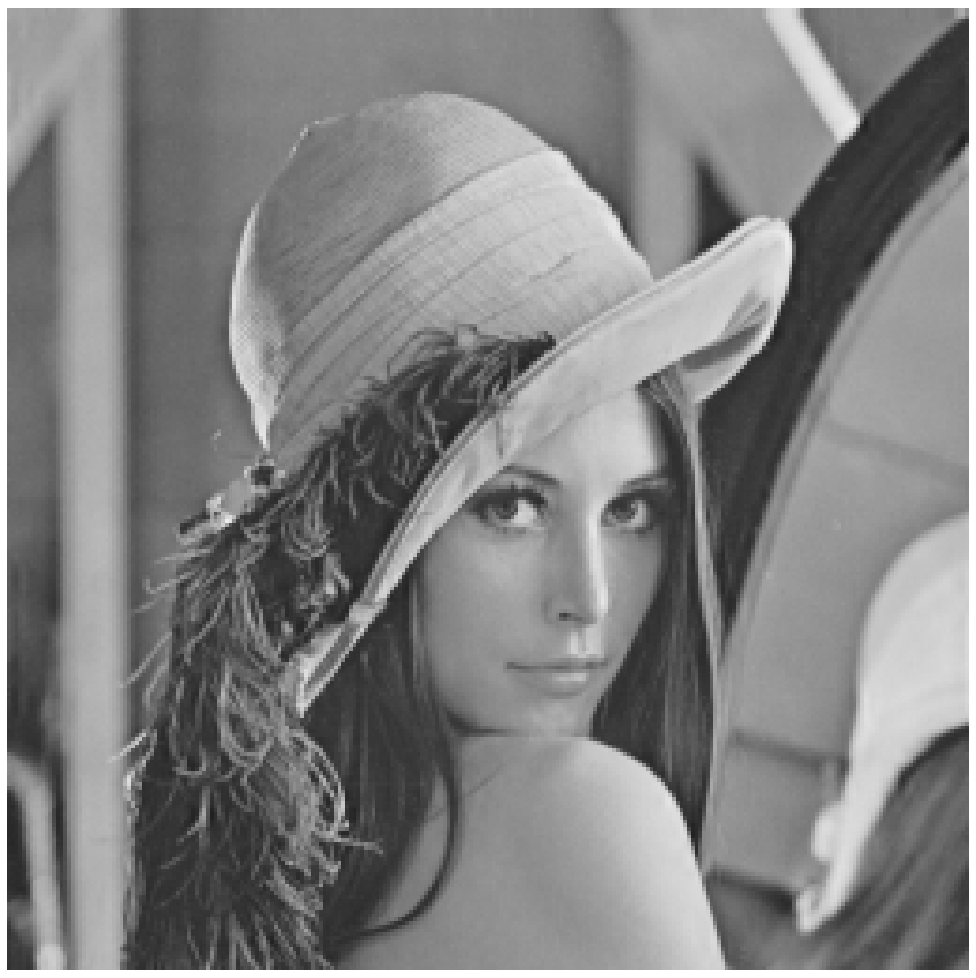

(b)

(c)

Fig. 7. Example of digital watermarking using visual watermarks, where (a) is the host *Lena* image, (b) is the watermarked image with *PSNR*= 45.68 after embedding, (c) the watermarks.

Table 2. The extracted watermarks from Fig. 7 are visually recognizable

| Embedded watermarks | NCU CSIE |
|---|---|
| *PSNR* (dB) after embedding watermarks | 45.68 |
| Extracted watermarks | NCU CSIE |
| Correlation | 1 |
| Error bits | 0 |

## 4.2. On the robustness against JPEG lossy compression

Table 3 shows the extracted results from *JPEG* compressed version of the watermarked images with different compression quality. The quality of watermarked images is still in good situation even under the high compressed ratio. The extracted watermarks and the original watermarks are with high correlation. In the proposed method, error rate of the extracted watermarks is rarely small (<0.5%) even under the situation of compression ratio, 16.59. From the experimental results, we can find that the proposed watermarking techniques yield satisfactory results in terms of transparency for watermarked images.

Table 3. Changes of correlation and error rate values of the proposed method under *JPEG* lossy compression.

| Attack | Extracted Watermark | Correlation | Error rate |
|---|---|---|---|
| Quality=100, CR=1.6810 | 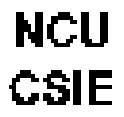 | 1 | 0 |
| Quality=90, CR=4.6681 | 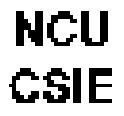 | 1 | 0 |
| Quality=80, CR=7.1466 | 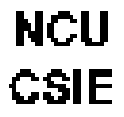 | 1 | 0 |
| Quality=70, CR=9.1314 | 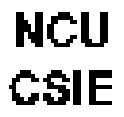 | 1 | 0 |
| Quality=60, CR=10.9222 | 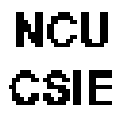 | 1 | 0 |
| Quality=50, CR=12.4298 | 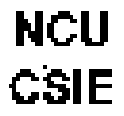 | 0.9972 | 0.09% |
| Quality=40, CR=14.1922 | 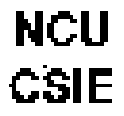 | 0.9972 | 0.09% |
| Quality=30, CR=16.5903 | 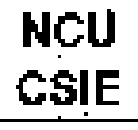 | 0.9864 | 0.49% |
| Quality=20, CR=20.5619 | 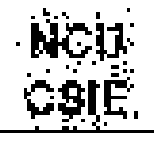 | 0.8022 | 7.8% |

*CR=compression ratio

## 4.2. On the robustness against JPEG lossy compression

Sharpen operations are used to enhance the subjective quality. Table 4 shows the extracted results of applying enhanced operation to a watermarked image. The extracted results are highly similar to the original watermark.

Smoothing operations such as median filter are used to decrease spurious effects that may be present in images from a poor transmission channel. Table 4 shows the extracted results of applying median filter to a watermarked image. The extracted watermark is still visually recognizable.

Table 4. The extracted of watermarks, correlation and error rate under sharpen and smoothing operation.

| Attack | Extracted Watermark | Correlation | Error rate |
|---|---|---|---|
| Sharpen | 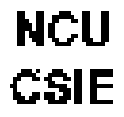 | 0.9945 | 0.19% |
| Median filter | 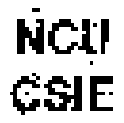 | 0.9488 | 1.8% |

## 5. CONCLUSION

We have introduced a watermarking framework for embedding visually recognizable watermarks in images, which can resist image-processing attacks, such as *JPEG* compression, smoothing, sharpening, blur, *etc*. The proposed new techniques are based on the context ideas and fuzzy filter. In the experiments, the fuzzy filter is employed to conclude the entropy of each coefficient in the center of the context. The watermark embedding strategies of the proposed methods consider the local characteristics by choosing the larger-entropy coefficients of *DWT* subbands to embed watermarks. The experimental results show that the proposed methods provide extra robustness against *JPEG*-compression and image processing compared to the traditional embedding methods. Moreover, the embedded coefficients in the proposed approaches have their own embedding degrees, which are calculated automatically in accordance with the entropy contributed from the context. Finally, the proposed approaches have no need the original host image to extract watermarks.

## ACKNOWLEDGEMENTS

The authors would like to thank Mr. Y.-H. Huang, Prof Tseng D-C, and the anonymous reviewers of this paper.